# Particle production in proton-proton collisions

M. T. Ghoneim, M. T. Hussein and F. H. Sawy
Physics Department, Faculty of Science, Cairo University, Cairo – Egypt
ghoneim@sci.cu.edu.eg, Tarek@Sci.cu.edu.eg and Fatma.Helal.Sawy@cern.ch

## *Abstract*

In this work, we present a study of particle production in proton-proton collisions using data that are collected from many experiments of relative wide range of reaction energies. These data include production of pions and heavier particles; like keons and lambda hyperons. Proton-proton collision is a simple system to investigate and to be considered a starting point that guides to more complicated processes of production in the proton-nucleus and the nucleus-nucleus collisions. In this paper, we are interested in the mechanisms that describe the process of particle production over a wide range of interaction energy, and how the physics of production changes with changing energy. Besides, this work may raise a question: are heavier particles than pions produced via the same mechanism(s) of producing pions, or these are created differently, being different in masses and other physical properties?

# I-Introduction

Particle production, the concept of mass-energy relationship, has been always one of the selected topics to investigate in high energy nuclear reactions over several decades. This type of research has probably started by the time people wanted to accelerate particles up to relativistic speeds and to smash them into other particles and see what may turn out.

Passing over the techniques of acceleration and particle detection, people observed that in p-p collisions at relativistic energy, more particles came out than those went in. The extra came out particles, as the acceleration of particles exceeded the threshold of particle production were, principally, created pions.

Providing the colliding protons with higher energy, particles would be created with higher multiplicity including heavier particles than pions. Of our tasks in this research is to follow up the variation of the multiplicity of created particles with the reaction energy by making use of the available reliable data over a relative wide range of laboratory energy; starting from few GeV up to several TeV. Creation of heavier particles than pions, is also considered wherever available.

## II-Particle creation between theory and experiment:

The nature of proton-proton interactions varies with energy as a result of the decrease of the strong coupling constant with energy as shown in figure (1). This fact could mean that there is no unique theory for describing particle creation mechanism in p-p interactions over the whole available energy range. It goes through phenomenological models in low energy region to perturbative quantum chromo-dynamics (PQCD) in high energy one [1-4]. The low energy collisions (low transverse momentum events or "soft events") lead to soft creation mechanism which is considered to be the basic mechanism for particle creation. For many years, a number of simplified phenomenological models have been developed in order to describe low energy production events; such as the string model in its simple form [5].

If one starts with the basic assumptions of the string model for the p-p collision, and considers that the p-p collision would go through binary interactions between their constituent quarks, the number of participant quarks in the collision depends upon the impact parameter according to geometrical aspects. The range of the impact parameter is probably divided into three regions: the most peripheral, the intermediate and the central ones, indicating one, two and three interacting pairs, respectively. In the second and third regions the interactions may go through either individual quark-quark (q-q) or collectively, i.e., more than one quark may behave as one entity. The energy would be distributed among the participant quarks. The string model, basically, assumes the presence of a string between the interacting quarks due to the color field that exists between them. It assumes also that hadrons are formed in two structure events by the fragmentation of the string. In other words, strings are formed by quark-quark, quark-diquark and by diquark- diquark systems, where a primary string is formed among the originally interacting quarks while a secondary string is generated in the field of sea quarks of the primary string and so on for higher order strings. A string continues producing particles in a ranking order governed by a scaling fragmentation function until the remnant energy is less than the threshold of production. The string force between the color charges causes the quarks to fly apart under the effect of its repulsive part, decelerate due its attractive part, accelerate back together and then fly apart once more, thus, executing periodical oscillations. The color force field may materialize a massless $q\bar{q}$ pair of zero energy-momentum at a point of the string. The string then separates into two independent color-neutral strings. As time develops, the string breaks randomly into smaller pieces carrying smaller fraction of the initial energy. When the mass of a string piece gets small enough, it is identified as a hadron and the breaking stops within that piece meaning that the whole system eventually evolves into hadrons. The average multiplicity of the "softly" created particles $\bar{n}$, shows a linear logarithmic relationship with the natural logarithm of the center of mass energy, s:

$$\bar{n} = 1.8 \ln s - 3.4 \ldots\ldots\ldots (1)$$

This linearity, described by the above equation, and figure (2), is supported by the Fermi scaling and also predicted by most of theoretical models like "multi-peripheral model" and "quark-parton model" in asymptotic soft regions many years ago[6,7]. The parameterization works reasonably and this linearity fits well up to the Intersecting Storage Rings (ISR) energy range.

Soft creation is also dealt with by the Dual Parton Model (DPM) that coupled unitarity with the parton structure of hadrons to describe soft collisions among protons by single pomeron exchange. The soft collisions separate the valence quarks of each incident proton into two colored systems a quark and a di-quark, as the string model proposed. Chains are assumed to be formed by pomeron exchange where each pomeron gives rise to two chains stretched between valence partons. Fragmentation of colored systems results in production of hadronic chains. A unitarity cut gives two chains of hadrons, and hence the leading contribution in soft creation is two chains that are stretched between only valence quarks of the initial hadrons [8].

The multiplicity distribution of created particles by soft events follows a Poisson distribution, meaning that every single final-state particle is created and emitted independently [9]. The emission of these created particles is viewed as a black body radiation, the statistical approach of which was pioneered, for more than 50 years ago, by Fermi. The KNO scaling of multiplicity distribution (blow ISR range) for non-single diffractive NSD events in full phase space supports a single creation mechanism [10]. So, the Poisson distribution of the multiplicity of created particles and their obeying of KNO scaling, support the thought that creation is a pure soft mechanism up to ISR range.

Above the ISR energy range, the multiplicity distribution showed a deviation from the Poisson shape that was referred to some kind of correlations between the created particles as well as a sign of variation in the creation mechanism. These correlations were proposed by the "Clan Model" which assumes the ability of a particle to emit additional particles, as cascading by decay and fragmentation. The model considers that the created particles stream as clans (clusters) where the ancestors production, and thus the clans, are governed by a Poisson distribution [11]. A clan contains all particles that stem from the same ancestor, where the ancestors themselves are produced independently. The interpretation of "Clan Model" was based on the success of the Negative Binomial Distribution NBD to describe multiplicity distributions up to √s = 540 GeV, announced by UA5. The same meaning could be found in the string model that proposed multi-order fragmentation for higher energy. The multi-order fragmentation is a cascading process of ordinary fragmentation due to high energetic partons limiting to reach the threshold energy for materialization into hadrons. Hadrons are formed in the successive fragmentation of the string, and this also means that as the reaction energy increases, the number of hadrons produced increases and some hadrons are correlated, not independently like the case of lower energies. The interpretation had got more support by assuming new type of events, called "semi-hard" events in addition to "soft" ones that produce these bundles. Experimentally, semi-hard events are responsible for a "mini-jet" production [9]. A mini-jet, according to UA1 collaboration, is a group of particles that have a total transverse momentum larger than 5 GeV/c [12, 13]. As energy goes higher than 540 GeV; semi-hard events start to show significant contribution in collision and many models were modified by adding new terms that represent them. Corresponding to DPM, mini-jets are generated from at least four chains, two of them come from contribution of valence quarks and other two are generated from sea partons through semi-hard interaction. As energy goes higher, sea parton contribution goes bigger. Since sea partons carry only a small fraction of the momentum of the incident hadrons, the chains are concentrated in the central rapidity region. Thus, these extra chains explain the rise of the central particle density, and DPM calculates this contribution of additional chains. Since this large rise in created particle multiplicity originates in the central region, KNO scaling has been violated in this region. This violation is traced to short range correlations of particles in the strings and interplay between the double-pomeron processes [9, 12]. The superposition of the two types of interaction affects the multiplicity distribution and therefore potentially explains the deviation from the scaling found at lower energies. UA5 has been successfully fitted multiplicity distributions of created particles as a superposition of two independent NBDs. This interpretation has been confirmed by the success of two NBDs at √s = 900 GeV up to √s = 1800 GeV, and the inability of single NBD to describe multiplicity distribution in that range of energy [9]. So, the possible scenario for particle creation up to √s = 1800 GeV is a weighted superposition mechanism that combines two classes of events for one and the same mechanism. Therefore, no interference terms have to be considered, and the final distribution is the sum of the two independent distributions, which is supported by a two-component model by Giovannini and Ugoccioni, in 1999, that divides particle creation mechanism into two-components in that energy range [13].

Thus, the total mean multiplicity $\bar{n}$ was divided also into two parts; one from a soft component $\bar{n}_{\text{soft}}$, and the other from a semi-hard component, $\bar{n}_{\text{semi-hard}}$:

$$\bar{n} = \bar{n}_{\text{soft}} + \bar{n}_{\text{semi-hard}} \quad \ldots\ldots\ldots\ldots\ldots\ldots\ldots (2)$$

That is why equation (1) could not describe the data up to 1.8 TeV, but adding a quadratic natural log-term to the linear part was found to show a better fit up to that limit, as expressed by the following equation and shown in fig.(3):

$$\bar{n} = 2.7 - 0.58 \ln s + 0.21 (\ln s)^2 \quad \ldots\ldots\ldots\ldots\ldots\ldots\ldots (3)$$

The quadratic term, in the above equation, reflects the contribution of semi-hard and gluon-bremsstrahlung process [13]. The gluon bremsstrahlung represents particle creation source that originates at TeV energy collision, commonly known as Initial and Final State Radiations, respectively. Initial and Finial State Radiation emit gluon before and after real collision have occurred, respectively. This radiation (gluon) can materialize to produce hadrons and the increase in collision energy, increases the contribution of this radiation in particle production.

As energy goes higher, the strong coupling constant $\alpha_S$ becomes small, resulting in asymptotic freedom. Collisions of p-p at this higher energy can be viewed as parton-parton collisions which can be mathematically described by perturbative quantum chromo-dynamics (PQCD). These high energy parton collisions generate new type of events called "hard" events in addition to "soft" and "semi-hard" ones. Hard parton interactions develop via short-distance over a very short time scale and the subsequent fragmentation produces a cone of hadronic final states that originate from the same partons. This cone of hadrons is called a "jet"; meaning that a jet production is an evidence for the existence of hard collision, with each jet representing an independent fireball for hadron creation. This picture suggests that the properties of the jet depend only on the initial parton. The proposal of DPM is forming more than four chains by multiple pomeron exchanges and makes the central plateau height increase with energy, therefore, the multi-chain contribution becomes increasingly important and the average number of chains increases with √s. From few GeV interaction energy up to √s = 7 TeV, the mean multiplicity was found to be best described by cubic logarithmic energy dependence, while quadratic and linear dependences succeeded to describe it at other lower parts of that range, as is given in fig.3 and the following equation:

$$\bar{n} = -4.0 \times 10^{-3}(\ln s)^3 + 0.28(\ln s)^2 - 0.86\ln(s) + 2.9 \quad \ldots\ldots\ldots\ldots\ldots\ldots\ldots\ldots(4)$$

The quadratic and cubic terms in the above phenomenological formula, which is already supported by some other researchers[14], were interpreted by another point of view stating that these terms come from double and triple parton interactions, respectively, rather than by only one [15,16]. This explanation might compete with the former one to stimulate the expectation of the presence of further higher order terms of "ln s" in the above formula as energy goes higher and higher in future experiments.

From our analysis of particle creation over investigated relative wide range of energy, so far; one may reach at the thought that the behavior of the soft interaction mechism is the dominant one over the stated range of energy and that the quadratic and cubic behavior are atributed to semi-hard (mini jet) and hard (jet) behavior, respectively. One may notice also that the quadratic and cubic curves do not go far from each other all over the stated range, as can be noticed in figure (3). This observation might be referred to the fact that the cubic term is weighed by a lower probability than the quadratic one; a matter of cross-section of occurrence.

### III-Heavier particle creation:

The above work, so far, has considered creation of particles without distinguishing between their entities. It was found interesting to consider the creation of different types of particles separately, the data of which were gathered from some references, wherever available [17-19]. Again, the relations between the mean multiplicity of created pions, keons and lambdas, $\bar{n}_\pi, \bar{n}_k, \bar{n}_\Lambda$, respectively; as a function of ln (s), were investigated and the fitting graphs were made to guide the eyes. The data fitting was found to give the following phenomenological equations:

$$\bar{n}_\pi = 0.02(\ln s)^3 - 0.20 (\ln s)^2 + 2.34 (\ln s) - 3.8 \quad \ldots\ldots\ldots\ldots (6)$$
$$\bar{n}_k = 0.001(\ln s)^3 - 0.01 (\ln s)^2 + 0.16(\ln s) - 0.37 \quad \ldots\ldots\ldots\ldots (7)$$
$$\bar{n}_\Lambda = 0.03 (\ln s) - 0.1 \quad \ldots\ldots\ldots\ldots (8)$$

The above fittings together with their experimental date are shown in figure (4).

From the above equations and figure [4], threshold energies for the creation, $E_{th}$, were estimated to give 6.5, 13.5, 28 GeV for pions, keons and lambdas, which correspond to incident lab momenta of $P_{lab}$=1.6, 5.5 and 12.9 GeV/c, respectively.

It was found worthy to look at the above given data from the point of view that considers the relative probability of production of each type of the considered particles with respect to the creation of pions ($R_{\pi/keon}$ and $R_{\pi/\Lambda}$) over the available energy range. These ratios are given in figure [5].

From equations (6-8) and figure [5], one may notice that below certain values of ln s, both probabilities are remarkably large, then they show sudden drop to much lower values and go almost steady beyond that value of ln s. one may notice also that the ratio $R_{\pi/\Lambda}$ is appreciably higher than $R_{\pi/keon}$ over all the available energy range.

These couple of notices may derive one to think about the practical creation threshold energy of pions, keons, lambdas and consequently of higher particles; to be looked upon as changes of the state in the fine structure of the nuclear matter. Such a change (or changes) allows the production of different particles according to the opening of some channels at some energy values according to some selection rules of creation of different particles with different probabilities. A very simple picture that could resemble our point is a mixture that is composed of several liquids, when heated, would evaporate each liquid at a time according to the "boiling point" of that liquid. That is why we believe, as many physicists do long ago, that higher energy would always lead to new physics.

The fact that creation of pions is more probable than creation of heavier particles was referred to the low probability of creating s quarks, which is responsible for the formation of heavier particles than pions in spite of the presence of a lot of energy. This thought assumes the credit of more production of lighter quarks like u and d, than of production of s ones, which in turn, was believed to come from the color field nature which loses its energy bit by bit as excitations (kinks) of soft gluons rather than by hard single gluon radiation [9, 20-22]. Moreover, it was added to the former explanation, the fact that pions are the too much more stable mesons, than the other particles which decay quickly to pions, and they also have very short life times. This interpretation may be fairly a satisfying physics to account for the differences in creation of different particles. The graphs of creation ratio show always a much higher values than 1, which really supports the priority for pion creation even at higher energies than the threshold for creation of heavier particles, but nobody can deny that the same graphs show that something is taking place in the constituents of the nuclear matter behavior at some energy values.  More sudden changes in that behavior are expected to show up in other experiments dealing with creation of heavier particles than pion, keon and lambda at higher energies.

## Conclusions

The physics of particle creation changes with changing energy and the mechanism could be divided into three parts: soft, semi-hard and hard components along the studied energy range. The soft component for particle creation mechanism (which means that the final-state particles are created and emitted uncorrelated) seems to be the dominant one allover the range of energy considered. However, it is, probably, the only mechanism in particle creation up to the limit of the ISR. This nature changes as energy goes higher, where particles have the ability to emit additional particles by decay and cascading production. This leads to the addition of a semi-hard component to the soft one to take part in the same mechanism. At the TeV region, a hard component takes a leading role in creation in addition to the other components.  Heavier particle production (like keons and lambdas) show remarkable lower cross-section than pion, the fact that was attributed to the small bits of energy the color field loses as excitation of soft gluon rather than by hard single gluon radiation. The results also show that creation of heavier particle than pions may require some kind of change in the state of the nuclear matter constituents. This might be thought of "boiling points" and "latent heats", somehow, to be necessary for creation of different particles from the nuclear matter.

### Acknowledgements
The authors are so thankful and grateful to prof. Daniel Denegri, senior research scientist and staff member at CERN, and prof. Paris Sphica from University of Athens for their fruitful discussions.

# References

[1] Daniel Stump, Joey Huston, Jon Pumplin and Wu-Ki Tung, *JHEP*, 10-46, 2003.
[2] S. S. Adler et al. (PHENIX Collaboration), *Phys. Rev. Lett.*, 91-241803, 2003.
[3] T. Regge. *Nuovo Cimento*, 14-951, 1959.
[4] V. N. Gribov, *Sov. Phys. JETP*, 53-654, 1967.
[5] M.T. Hussein, A. Rabea, A. El-Naghy and N. M. Hassan, progress of theoretical physics, vol.93, No.3, 1995.
[6] V. Cerny and J. Pisut, Acta Physica Polonica Vol. B8, No.6, 1977.
[7] M. El-Nadi, A.T. Baranik, A. El-Naghy, N. Mettwelli, S. Abd El-Halim and F. Ahmed, Egyptian Journal of physics, Vol. 2, No. 14, 1983.
[8] Uday P. Sukhatme, XXII international symposium on multi-particle dynamics, Santiago De Compostela, Spain, July 1992.
[9] Niccol`o Moggi, faculty of mathematical, Physical and Natural Sciences, university of Pavia, Italy, Ph.D thesis in Physics-XI cycle,1998/1999
[10] Z. Koba, H.B. Nielsen, P. Olsen, Nucl. Phys. B40 (1972), 317.
[11] Jan Fiete Grosse-Oetringhaus and Klaus Reyger, J. Phys. G: Nucl. Part. Phys. vol. 37, no. 8, 2010.
[12] J.Tran Thanh Van, Current Issues in Hadron Physics: Proceedings of the XXIIIrd Rencontre De Moriond, Les Arcs, Savoie, France, 1988.
[13] A.Giovannini and R.Ugoccioni, Phys. Rev. D60 (1999) 074027.
[14] Ashwini Kumar, B. K. Singh, P. K. Srivastava, C. P. Singh, The European Physical Journal Plus (2013) 45-128.
[15] T. Alexopoulos, E.W. Anderson, N.N. Biswas, A. Bujak, et al., Phys. Lett. B435 (1998) 453-457.
[16] W. D. Walker, Phys. Rev. D69 (2004) 034007.
[17] M. Antinucci, lettere El-Nuovo Cimento vol. 6, no. 4, 1973.
[18] T. Anticic, Eur. Phys. J. C (2010) 68: 1–73
[19] R. E. Ansorge, Nuclear Physics B328 (1989)36-58.
[20] Tai An and Sa Ben-Hao, Phys. Rev. C 57(1998) 261.
[21] Walter Greiner, Horst Stöcker and André Gallmann, Hot and Dense Nuclear Matter, Plenum press - New York (1994).
[22] R. K. Ellis, W. J. Stirling, B. R. Webber, QCD and Collider Physics, the press syndicate of the University of Cambridge (1996).

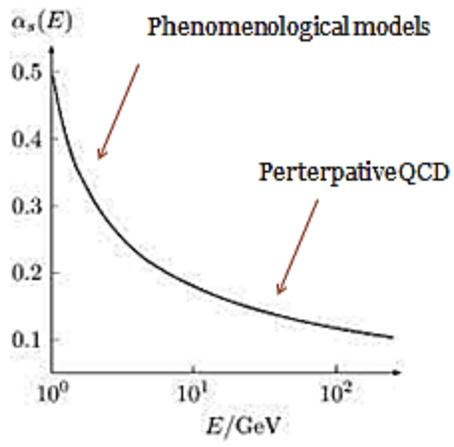

Fig. 1

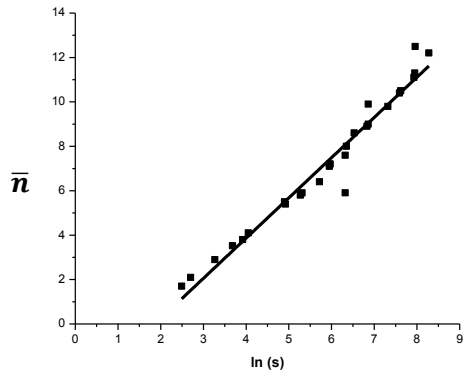

Fig. 2

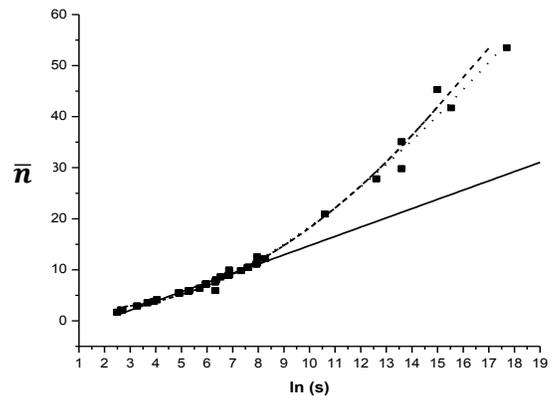

Fig.3

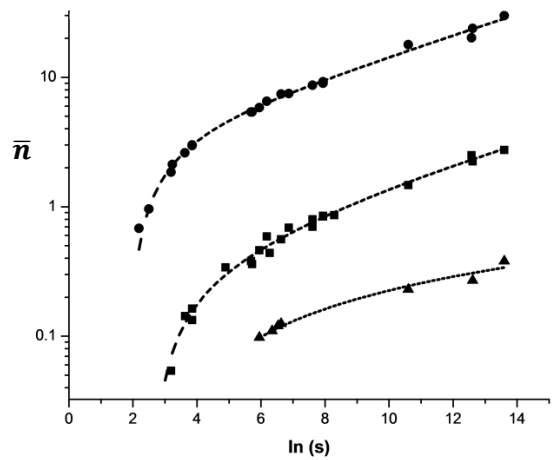

Fig.4

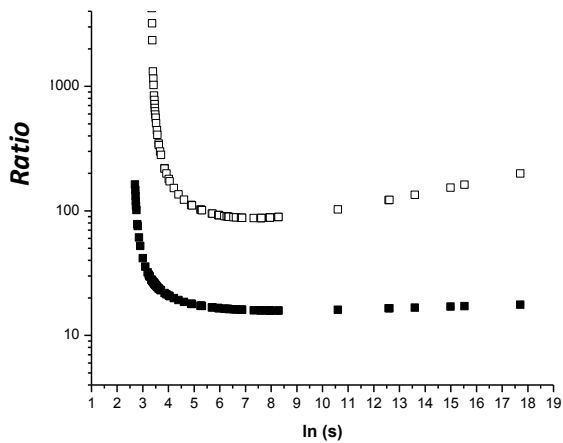

Fig.5

# Figure captions

Fig (1): the relation between the strong coupling constant with energy.

Fig (2): the mean multiplicity of created particles, $\bar{n}$, as a function of ln(s) up to ISR energy limit. Points are experimental data and the straight line is the fitting relation of equation (1).

Fig (3): the mean multiplicity of created particles, $\bar{n}$, as a function of ln(s). Points are experimental data for NSD events in full phase space of *p-p and/or* $p-\bar{p}$ collisions as a function of ln(s) up to √s = 7 TeV. Solid line, dashed and dotted curves are the fittings of equations (1), (3) and (4), respectively.

Fig (4): the mean multiplicities of π's, k's and Λ's as a function of ln(s): asterisks, squares and triangles are experimental data, while solid, dashed and dotted lines are their corresponding fittings, respectively.

Fig (5): the dependence of the ratio of mean multiplicities: pion to keon $R_{\pi/keon}$ (empty squares), and pion to lambda $R_{\pi/\Lambda}$ (solid squares) on ln(s).